\documentclass[runningheads]{llncs}
\usepackage{hyperref}
\usepackage{graphicx}
\begin{document}
\title{One-shot screening of potential peptide ligands on HR1 domain in COVID-19 glycosylated spike (S) protein with deep siamese network}
%
%
\author{Nicol\`o Savioli}

%

\institute{Imperial College London \\
\email{n.savioli@imperial.ac.uk}}

\maketitle              
\begin{abstract}
The novel coronavirus (2019-nCoV) has been declared to be a new international health emergence and no specific drug has been yet identified.
Several methods are currently being evaluated such as protease and glycosylated spike (S) protein inhibitors, that outlines the main fusion site among coronavirus and host cells. 
Notwithstanding, the Heptad Repeat 1 (HR1) domain on the glycosylated spike (S) protein is the region with less mutability and then the most encouraging target for new inhibitors drugs. 
The novelty of the proposed approach, compared to others, lies in a precise training of a deep neural network toward the 2019-nCoV virus.
Where a Siamese Neural Network (SNN) has been trained to distingue the whole 2019-nCoV protein sequence amongst two different viruses family such as HIV-1 and Ebola.
In this way, the present deep learning system has precise knowledge of peptide linkage among 2019-nCoV protein structure and differently, of other works, is not trivially trained on public datasets that have not been provided any ligand-peptide information for 2019-nCoV. 
Suddenly, the SNN shows a sensitivity of $83\%$ of peptide affinity classification, where $3027$ peptides on SATPdb bank have been tested towards the specific region HR1 of 2019-nCoV exhibiting an affinity of $93\%$ for the peptidyl-prolyl cis-trans isomerase (PPIase) peptide. 
This affinity between PPIase and HR1 can open new horizons of research since several scientific papers have already shown that CsA immunosuppression drug, a main inhibitor of PPIase,  suppress the reproduction of different CoV virus included SARS-CoV and MERS-CoV. Finally, to ensure the scientific reproducibility, code and data have been made public at the following link: \url{https://github.com/bionick87/2019-nCoV}.
\end{abstract}

\section{Introduction}

The 2019-nCoV has risen in the city of  Wuhan in China’s Hube as an extraordinary infected human pathogen, producing critical respiratory illness and pneumonia. 
On 30 December 2019, three bronchoalveolar lavage samples were obtained from a patient with pneumonia of unfamiliar etiology. The complete virus genome sequence of 2019-nCoV was acquired and indicated a relationship bat SARS-like coronavirus \cite{WuF}.

At present, numerous coronavirus sequences have been published on GenBank \cite{Sayers}, allowing a direct study of the virus structure that the high mutation and recombination rates of the virus make it difficult to design a wide spectrum inhibitor at a conventional targets \cite{Xue,Yang}.

Several approaches are being evaluated. The first approach is the virus protease inhibition that prevents the virus from polypeptide filaments to be split and therefore the viral core proteins cannot be built.
For instance, Nelfinavir, an HIV-1 protease inhibitor to treat HIV, was prophesied to be a likely inhibitor of 2019-nCoV principal protease by different molecular docking computational-based \cite{Xu}.
The second strategy is targeting the glycosylated spike (S) protein in the fusion with the entry of the host cell \cite{Wrapp1}, which represents the most promising target for developing new inhibitors for the target site spike (S) S-HR1 \cite{Liu,Xia1}. Notwithstanding, it is unclear whether 2019-nCoV also holds a similar fusion and entry mechanism as that of SARS-CoV and MERS-CoV, and if true, consequently, the S-HR1 site can also serve as an important target for the development of 2019-nCoV fusion/entry inhibitors. Though, it is also unexplored if 2019-nCoV also exists other comparable fusion and entry mechanisms with SARS-CoV and MERS-CoV \cite{Xia2}.

Despite this, all of these deep learning methods are sharpening on the engendering of new molecules that have not yet been clinically tested.
Conversely, computational drug repurposing gives an efficient and fast approach to test drugs already available \cite{Karaman}.
Among this repurposing computational approach, four molecules have previously been selected to be the main candidates: Prulifloxacin, Bictegravir, Nelfinavir, and Tegobuvi \cite{Li}. Where Prulifloxacin is a synthetic antibiotic of the fluoroquinolone class \cite{Nelson}, Bictegravir is and an antiretroviral to block the enzyme integrase, used in HIV-1, capable of inserting a viral genome into a host one \cite{Tsiang}, Nelfinavir is protease inhibitors well used in the treatment of HIV-1 \cite{Zhang2}, Tegobuvi is a non-Nucleoside Reverse Transcriptase Inhibitor (NNRTI)  with exhibited antiviral activity in patients with genotype 1 chronic HCV infection \cite{Hebner}. 

However, none of these molecules has any specific inhibition targeting for the glycosylated spike (S) proteins.
Specifically, protein spike (S) is made up of two subunits: S1 and S2. The S1 subunit binds the cell receptor with its receptor-binding domain (RBD), followed by a set of conformation changes in the S2 subunit, allowing the fusion peptide to enter the cell membrane of the host cell.
In the S2 region, we find another region called heptad repeat 1 (HR1) with three hydrophobic grooves that bind to a high region called heptad repeat 2 (HR2) forming a structure with six helices (6-HB), which helps to bring the two membranes closer together for the final fusion and hence the entrance. The RBD of any CoV family is a highly mutable region and not suitable for inhibition, while the HR region in the S2 subunit is conserved among various HCoVs \cite{Xia1}.

Recently, several deep learning algorithms have been employed for the research of coronavirus target drugs. 
For example, three-dimensional modeling of the virus where 2019-nCoV sequences are translated into protein and then, within a classification network among ligand and protein as input, their interaction is screened \cite{Zhang,Rishikesh,Markus}. While generative adversarial approaches were also used for producing novel target molecular structures \cite{Zhavoronkov,Polykovskiy,Gao,Vijil}.
However, the two main limitations of these approaches are: (i) they are mostly with supervised training approach on public datasets in which there is not any specific ground truth ligand for 2019-nCoV (ii) they are based on shallow and non-convolutional networks (currently the state of the art in many datasets), (iii) they generate new molecules that have not yet been tested clinically.
 
For all these reasons, the purposes of this paper are: (i) Introduce a new artificial intelligence model that speeds up the research for the promising target ligand and is not based on complex and computationally expensive molecular docking operations. (ii) Fast screening of the main antibacterial, anticancer and antimicrobial peptides present in SATPdb \cite{Sandeep} toward the HR1 domain which that, as already mentioned, describes the more limited mutation spike protein site of 2019-nCoV. (ii) Moreover, introduce a new deep learning method, that is not based on a trivial supervised classification of ligand and peptide on public datasets, but on one-shot learning approach for specifically studying the glycosylated spike (S) 2019-nCoV protein.

\section{Proposed analysis work-flow}

The work-flow is subdivided into three stages: virus genome conversion into protein and subsequent splitting of the protein sequence within peptide filaments, text filaments peptide to image conversion, and finally the peptide comparisons with a Siamese Neural Network (SNN).

\subsection{Virus genome to protein}

The available genome GenBank 2019-nCoV sequence is used. The sequence was obtained from a 41-year-old man hospitalized in the Central Hospital of Wuhan on 26 December 2019. Each viral genome structure was determined by the alignment sequence of two characteristic portions of the Betacoronavirus family: a coronavirus linked with humans (SARS-CoV Tor2, GenBank accession number AY274119) and a coronavirus correlated with bats (bat SL-CoVZC45, GenBank accession number MG772933) \cite{WuF}. The genomic sequence was then converted toward the corresponding protein, where a sequence of ten protein was extracted in series for forming a protein peptide (Fig. 1).

\begin{figure}[ht]
  \centering
  \includegraphics[width=0.6\textwidth]{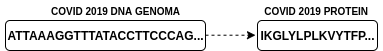}
    \label{fig:fig1}
  \caption{The figure shows the initial step of the workflow, that is, the translation of viral DNA into protein.}
\end{figure}

\subsection{Text peptide to image}

Subsequently, the protein-peptide is converted into an image of  $256 \times 256$ pixels (Fig. 2). This method is similar to DeepVariant work where the genomics sequence is transformed into an RGB pixel image and then directly processed by a state-of-art Convolutions Neural Network (CNN) for genotype prediction \cite{Poplin}. This enables handle single text strings as images and to use all the advantages of CNN that currently represent the state of the art in multiple tasks on several datasets \cite{Asifullah}. In particular, the smallest variations in the protein sequence can be efficiently recognized by the multiple nonlinear layers of the neural network. 

\begin{figure}[ht]
  \centering
  \includegraphics[width=0.3\textwidth]{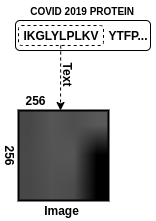}
    \label{fig:fig2}
  \caption{The figure shows the second step of the workflow, the translation of ten text amino acids to image.}
\end{figure}

\subsection{Peptide comparisons with a Siamese neural network}

A SNN \cite{Koch} is then trained to identify the whole 2019-nCoV protein structure versus two distinct family viruses such as Ebola \cite{Volchkov} and HIV-1 \cite{Frank}; where the equivalent genome-to-protein translation has been done. 
Though, this one-shot learning system reduces the use of specific datasets, enabling directly work on the available biological data. In other words, the network learns to discriminate target examples on its protein domain from other domains with completely different biological characteristics without the explicit use of a specific dataset.
Further is then applied deep SNN which i) lean generic image functionality to make inference on unknown distributions. ii) provide a valid approach that explores solutions on the unknown domain without relying on a trivial supervised training dataset (i.e. peptide ligand correspondence). iii) use several states-of-the-art  CNN previously pre-trained on huge datasets.

\begin{figure}[ht]
  \centering
  \includegraphics[width=1.1\textwidth]{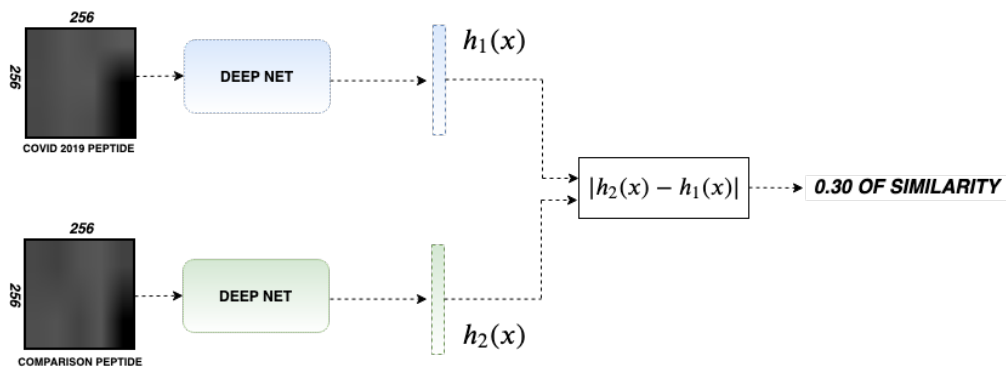}
      \label{fig:fig3}
   \caption{The figure shows the Siamese Neural Network (SNN) model.}    
\end{figure}

The model is then designed with two CNN that take distinct image inputs (Fig. 3). The input of the first network is always a sequence of 2019-nCoV peptide, while the second, can accept either a peptide sequence, randomly choose from different virus families (such as Ebola and HIV-1), or the same 2019-nCoV sequence (i.e as the first network). Therefore, if the two inputs are equivalent (i.e equal 2019-nCoV protein sequence) the ground truth target utilized for training the model is one; otherwise is zero.

Even though, two distinct types of CNN have used: a shallow and a deep version. Notably, for the shallow variant, AlexNet \cite{Krizhevsky} is employed. Especially, a  pre-trained version of AlexNet, on more than a million images from the ImageNet \cite{Deng} database, is also applied. While for the deep version is used the ResNeXt \cite{Saining} that is fifty layers deep.
Notably, the final convolutional layer of both CNN  is flattened into a vector and pair passed to a final layer that calculates their L1 metric within sigmoid output function.
The model is trained with Stochastic Gradient Descent (SDG) with Nesterov accelerated gradient (NAG) and learning rate $1e-4$ in $1000$ epochs where the loss function is mean squared error (MSE). While random resized crop, random horizontal flip, and random vertical flip are made to prevent overfitting during the training phase.

\section{Results}

To train the model, a dataset of $1258$ images is generated with the corresponding positive proteic example of 2019-nCoV \cite{WuF} and those negative ones of Ebola \cite{Volchkov} and HIV-1 \cite{Frank}. The dataset was subsequently divided into 60\% train, 20\% valid and 20\% test respectively. The model capacity to distinguish 2019-nCoV protein sequences from those of other viral families is estimated within a sensitivity analysis (Eq \ref{eq:1}).

\begin{equation}
Sensitivity  = 100*\frac{TP}{TP+FN}
\label{eq:1}
\end{equation}

Where $TP$ are true positive (i.e 2019-nCoV sequence correctly identified as 2019-nCoV sequence) while $FN$ are the false negative (i.e 2019-nCoV sequence incorrectly identified as Ebola or HIV-1).

\begin{figure}[h!]
  \centering
  \includegraphics[width=0.9\textwidth]{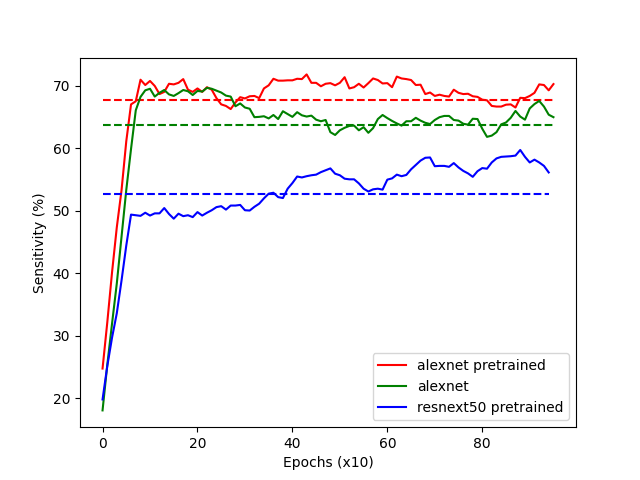}
      \label{fig:fig4}
   \caption{Validation plot during the training phase, as we can see AlexNet pre-trained on Imagenet \cite{Deng} shows a better ability to better learn from the earliest epochs compared to the not pre-perinated version. Instead, the deep network ResNeXt shows a more limited ability to adapt and learn; probably due to the small dataset available.}    
\end{figure}

As previously mentioned, two different CNNs were applied (i.e AlexNet and ResNeXt) for getting the best performance on the validation and test data set.
Importantly, the small eight layers AlexNet has been shown to perform in several tiny clinical datasets \cite{Savioli} and confirmed here to be the best network compared to a deeper version ResNeXt, also pre-trained on Imagenet  \cite{Deng}, made up by fifty layers.
Indeed, as pointed in the validation plot (Fig. 4), the pre-trained AlexNet shows a more reliable convergence correlated to the not pre-trained one and towards ResNeXt; this is probably due to the presence of small dataset to perform with large deep networks, that instead show slow convergence (i.e blu line Fig 4).
The sensitivity analysis on the test set confirms what was observed during the validation phase; whereas the pre-trained AlexNet version has more prominent sensitivity compared with its not pre-trained version and ResNeXt.

\begin{table}[]
\centering
\begin{tabular}{|l|l|l|}
\hline
\textbf{Models}                    & \textbf{Sensitivity}    \\ \hline
\textbf{AlexNet (pretrained)}      & \textbf{83.29 (2.55)}  \\ \hline
\textbf{AlexNet}                   & \textbf{70.56 (3.38)} \\ \hline
\textbf{Resnext-50 (pretrained)}  & \textbf{54.20 (3.43)} \\ \hline
\end{tabular}
\vspace*{5pt}
\caption{The table shows the sensitivity of the AlexNet (pre-trained and not) model respect with a more deeper model as Resnext in terms of mean and standard deviation.}
\label{tab:my-table}
\end{table}

After this deep network selection phase, the AlexNet pre-trained model is chosen to make an inference on $3027$ peptides from SATPdb \cite{Sandeep} to find the closest peptide to targeting HR1 2019-nCoV domain site. 

The inference process is performed as follows: (i) The HR1 sequence is cutting in eight sub-peptide of ten protein (Fig. 1) and then each converted into an image (Fig. 2) of $256 \times 256$ pixels. (ii) Separate SATPdb peptide is further converted from text to an image of $256 \times 256$ pixels (Fig. 2). (iii) Every single peptide is given in input to the first Siamese CNN network, while any of the eight  HR1 sub-peptides is paid to the second Siamese CNN sequentially. iv) Finally, the final Siamese sigmoid outputs, between the target SATPdb peptide with each of the eight sub-peptides, is averaged.
This inference process shows 93\% affinity between $IKKTYEEIKKTYEEIKKTYEEIKKTYEEIERDWEMV$ (peptidyl-prolyl cis-trans isomerase) peptide \cite{adam} and the HR1 domain.

\section{Discussion and conclusion}

In this article, a novel drug repurposing approach has been explored, where an SNN has been designed to find the best matching peptide inside the HR1 region; that represents the less mutability domain of any coronavirus \cite{Liu,Xia1}.
The proposed model achieves reasonable performance with an overall sensitivity on a test set of 83.29\% within two comparison models.
Particularly, a shallow pre-trained version as AlexNet has been applied showed better performance than a deep version one as ResNeXt (Tab. 1). 
An protein sequence dataset of $1258$ images was extracted from 2019-nCoV \cite{WuF},  Ebola \cite{Volchkov} and HIV-1 \cite{Frank} training the SNN to classify any protein sequences, similar to 2019-nCoV, compared to other belonging to different viruses family (i.e HIV-1, Ebola). 
Whereas, the lack of the training set size, justifies the greater ability of shallow networks to perform better compared to the more deep one (i.e ResNet).
However, the one-shot learning approach applied here is more suitable than a supervised ligand-protein classification \cite{Zhang,Rishikesh,Markus} because: (i) we do not need large train datasets to learn the correct ligands for 2019-nCoV, (ii) the neural network focuses the learning only in the interest 2019-nCoV protein sequence, (iii) public datasets have no training examples of ligand binding for the specific carnivorous 2019-nCoV (i.e which can create an incorrect classification).

The inference on $3027$ peptides on SATPdb \cite{Sandeep} had been taken, showing the confidence of 93\% for peptidyl-prolyl cis-trans isomerase affinity with  the HR1 2019-nCoV domain. The peptidyl-prolyl cis-trans isomerase (further known as peptidylprolyl isomerase or PPIase)  is an enzyme (Fig 5) that transforms the cis and trans isomers with the amino acid proline \cite{Lang}. The two major families of PPIase are Cyps and the FK506-binding proteins (FKBPs) where the Cyps are implicated in a broad spectrum of cellular processes including cell signaling, protein folding, and protein trafficking \cite{Naoumov}. The affinity between HR1 and PPIase may play an essential role in the identification of  PPIase inhibitors as several therapies with CsA immunosuppression drugs \cite{Frausto,Hopkins}. 
Several scientific papers have noted that, in cellular culture, the reproduction of different CoV (included SARS-CoV, and MERS-CoV) can be inhibited by CsA therapy \cite{Wilde,Pfefferle,Tanaka}. A well recognized CsA immunosuppression drug is the Sirolimus a PPIase inhibitor that has been shown to enhance results of patients with intractable H1N1 pneumonia \cite{Wang}, where recent computational studies show how Sirolimus can be also selected as a potential drug for 2019-nCoV \cite{Zhou}; confirming what has been observed in this work on HR1 and PPIase affinity.
In order to provide the reproducibility of this work, the code and data has been made available to the scientific community at \url{https://github.com/bionick87/2019-nCoV}.
Future work is thus still needed to confirm the association mechanism between HR1 and PPIase through molecular docking studies, therefore important for understanding biological mechanisms of PPIase on 2019-nCoV.

\begin{figure}[h!]
  \centering
  \includegraphics[width=1.1\textwidth]{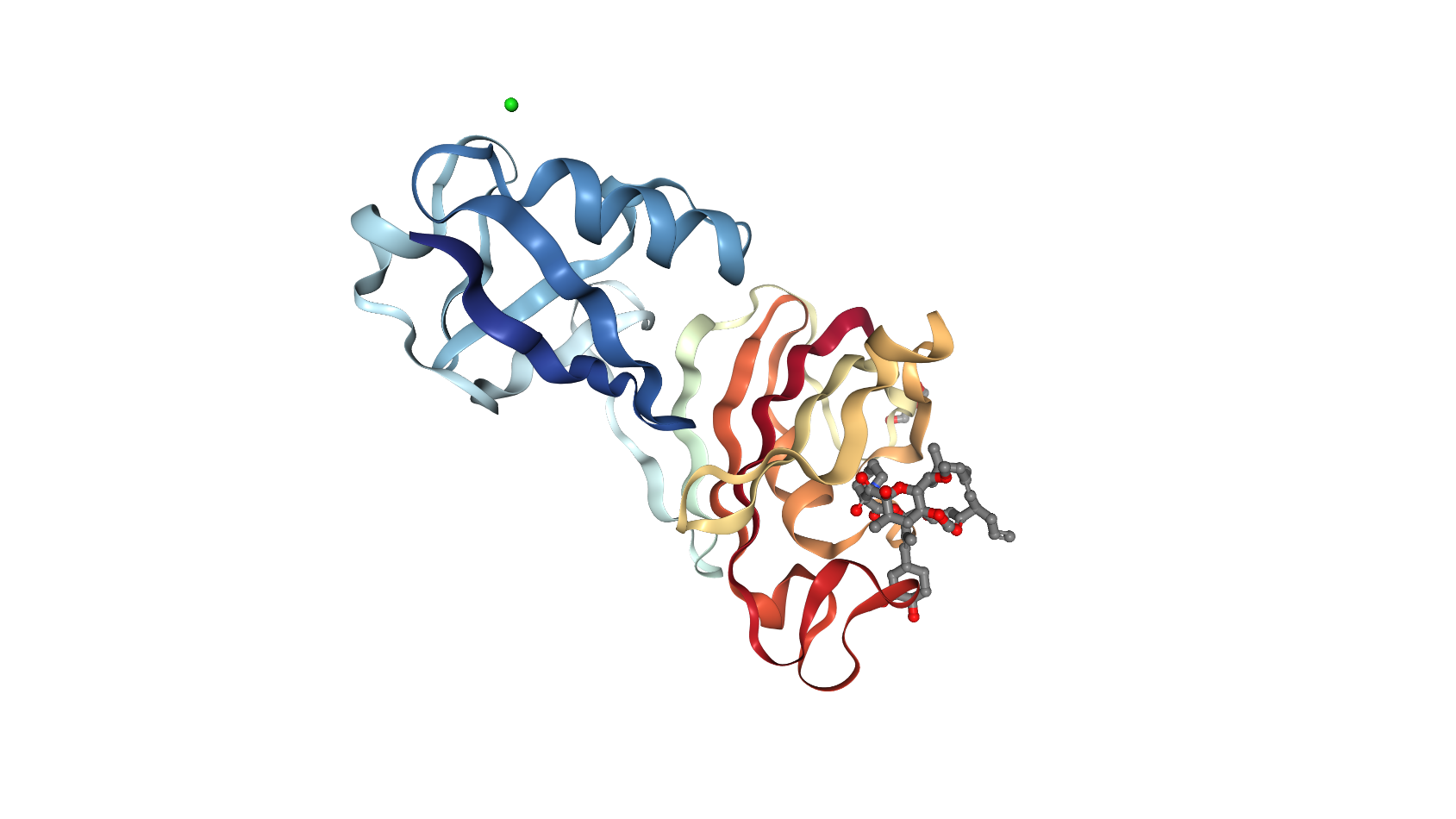}
      \label{fig:fig5}
   \caption{Peptidyl-prolyl cis-trans isomerase, images created with the PDB ID and associated publication, NGL Viewer (AS Rose et al. (2018) NGL viewer: web-based molecular graphics for large complexes. Bioinformatics doi:10.1093/bioinformatics/bty419), and RCSB PDB.}    
\end{figure}

%
%
%
%

\end{document}